\documentclass[12pt,3p]{elsarticle}

\usepackage{graphicx}               
\usepackage[caption=false]{subfig}  
\usepackage{amsmath, amssymb}       
\usepackage{wasysym}
\usepackage{lmodern}                
\usepackage[T1]{fontenc}            
\usepackage[utf8]{inputenc}         
\usepackage{physics}
\usepackage{caption}
\usepackage{subcaption}
\usepackage{natbib}

\begin{document}

\begin{frontmatter}

\title{Basin entropy and the impact of the escape positioning in an open area-preserving map}
\author[UFPR,URJC]{P. Haerter\corref{teste}}
\cortext[teste]{haerter@ufpr.br}
\author[UFPR,UFPR2]{R.L. Viana}
\author[URJC]{M.A.F. Sanjuán}

\affiliation[UFPR]{organization={Departamento de Física, Universidade Federal do Paraná},
            city={Curitiba},
            postcode={81531-990}, 
            state={Paraná},
            country={Brazil}}
      
\affiliation[UFPR2]{organization={Centro Interdisciplinar de Ciência, Tecnologia e Inovação, Núcleo de Modelagem e Computação Científica, Universidade Federal do Paraná},
            city={Curitiba},
            postcode={81531-990}, 
            state={Paraná},
            country={Brazil}}

\affiliation[URJC]{organization={Nonlinear Dynamics, Chaos, and Complex Systems Group, Departamento de Física, Universidad Rey Juan Carlos},
            city={Móstoles},
            postcode={28933}, 
            state={Madrid},
            country={Spain}}

\begin{abstract}
Efficient prediction of high-energy particle escape in toroidal fusion devices remains challenging due to the nonlinear dynamics of chaotic magnetic field line trajectories and their extreme sensitivity to initial conditions. Existing approaches vary from comprehensive magnetohydrodynamic models to simplified Hamiltonian mappings. We bridge this gap by employing a Hamiltonian mapping model of tokamak field lines featuring reversed shear profiles, combined with basin entropy analysis to quantify uncertainty in particle escape locations. Numerical simulations show that strategically positioning escape exits delays the onset of basin entropy growth, shifting critical entropy increases to higher levels of external perturbation. Our findings illustrate how nonlinear phase space structures can inform the design of plasma-facing components in fusion reactors, significantly improving predictive control of chaotic transport.
\end{abstract}

\end{frontmatter}

\section{Introduction}

Understanding the escape dynamics of high-energy particles in toroidal fusion devices, such as tokamaks, is critical for effective plasma confinement. However, the inherent nonlinearity of magnetic field line trajectories, which frequently exhibit chaotic and stochastic behavior, presents significant challenges
\cite{walkden2022physics,Krasheninnikov_Dynamics_2005,souza_Chaotic_2024}. These dynamics can be modeled by Hamiltonian systems, in which energy conservation precludes the existence of traditional attractors. Instead, escape basins serve as fundamental tools for analyzing particle transport. Although basins of attraction are well-studied in dissipative systems \cite{lichtenberg_Regular_1992}, their counterparts in conservative systems remain relatively underexplored.

Given a dynamical system, we define a basin of attraction as the set of initial conditions leading to a specific attractor. Hamiltonian dynamical systems, however, do not possess attractors, and consequently, do not have basins of attraction. Moreover, their trajectories can either remain bounded within a closed region in phase space or become unbounded and escape to infinity. When trajectories can escape, we classify the Hamiltonian system as open. In open Hamiltonian systems, we can define an exit or escape basin as the region in phase space containing initial conditions that lead trajectories toward a particular exit. A paradigmatic example illustrating this behavior is provided by the Hénon–Heiles Hamiltonian, in which the system transitions from bounded to open dynamics once the energy surpasses a certain threshold value, commonly termed the escape energy\cite{henon1964applicability}. Above this critical energy, the potential opens up, creating three natural exits through which trajectories may escape to infinity. The same concept can be extended to discrete dynamical systems, also known as area-preserving maps, or Hamiltonian maps, where instead of energy area is conserved. In these systems, the openings can even be introduced artificially, giving rise to various possible configurations\cite{altmann2013leaking}. In these systems, the precise placement of escape regions significantly influences the structure and characteristics of the resulting escape basins \cite{sanjuan_opening_2003, schneider_dynamics_2002, aguirre_wada_2001}.

Previous studies have utilized Hamiltonian maps to analyze chaotic magnetic field-line behavior and particle escape in tokamak devices. However, these works frequently oversimplify the escape mechanisms by defining exit regions arbitrarily, without a physically justified criterion \cite{mathias_fractal_2023, souza_fractal_2023, souza_Chaotic_2024}. Specifically, many prior analyses assume escape regions to be homogeneous or position them exclusively near hyperbolic regions in phase space. Such simplifications neglect the complex interplay between nonlinear dynamics and realistic physical boundaries, potentially leading to an incomplete understanding of how escape region placement influences particle transport and basin structures. 

The present work investigates how the selection and placement of escape regions affect both the structure of escape basins and basin entropy, highlighting the sensitivity of conservative dynamical systems to the configuration of their escape mechanisms. We address these issues through numerical simulations using the Revtokamak model, an area-preserving map designed to represent magnetic field lines within toroidal confinement devices, commonly used in plasma fusion experiments \cite{balescu_hamiltonian_1998}. The Revtokamak model was specifically chosen for its inclusion of reversed shear dynamics, a key feature in advanced fusion reactors known to significantly reduce turbulent particle transport, thereby enhancing plasma confinement\cite{panta2022control}. In contrast to previous studies, which often relied on arbitrarily or simplistically placed escape regions, we systematically vary exit locations within our model. This approach enables us to identify optimal configurations of divertor plates, with the ultimate aim of minimizing chaotic particle transport and thus contributing to improved plasma confinement strategies\cite{loarte2001effects}.

In this context, basin entropy plays a crucial role as a quantitative measure for analyzing the complexity and organization of escape basins in the Revtokamak model. Basin entropy, which uses Shannon entropy to quantify uncertainty in escape dynamics, has been particularly effective for detecting subtle structural transitions in basin topology across various dynamical systems. It has been demonstrated to be more sensitive to changes in escape basin structure than traditional nonlinear metrics, such as fractal dimension and the uncertainty exponent \cite{daza_basin_2016,daza_pt_2024}. By applying basin entropy to the Revtokamak model, we can precisely assess how variations in escape region placement influence the dynamical behavior of magnetic field lines, enabling more informed optimization of divertor configurations and better control of chaotic particle transport in fusion devices.

The motivation for studying the Revtokamak model arises from the critical need to better understand the escape of high-energy particles from plasma confinement regions\cite{zweben2011rapidly,ivanova2023divertor}. Such escapes pose significant challenges in tokamaks, as escaping particles can damage plasma-facing components and disrupt confinement, thus threatening overall plasma stability and reactor performance \cite{mathias_fractal_2022}. By systematically introducing and varying escape regions within this model, we can investigate precisely how particle trajectories exit the system and analyze the resulting structural evolution of escape basins under diverse dynamical conditions. Our results demonstrate that strategically positioning escape regions induces abrupt phase transitions in basin entropy, driven specifically by anticlockwise shifts in particle trajectories as external perturbations increase. This insight provides a deeper understanding of the relationship between escape basin complexity and escape mechanism placement, enabling improved predictive control of chaotic transport in fusion plasmas.

The paper is structured as follows: In Section \ref{sec:Tokamap}, we describe the Revtokamak model and the numerical methods employed in our analysis. Section \ref{sec:pri} investigates how the positioning of escape regions influences basin entropy and the complexity of basin boundaries. Finally, Section \ref{sec:concl} summarizes the key findings and presents the conclusions of this study.

\section{The Revtokamap Model}
\label{sec:Tokamap}

\begin{figure}[h!]
    \centering
    \includegraphics[scale=0.13]{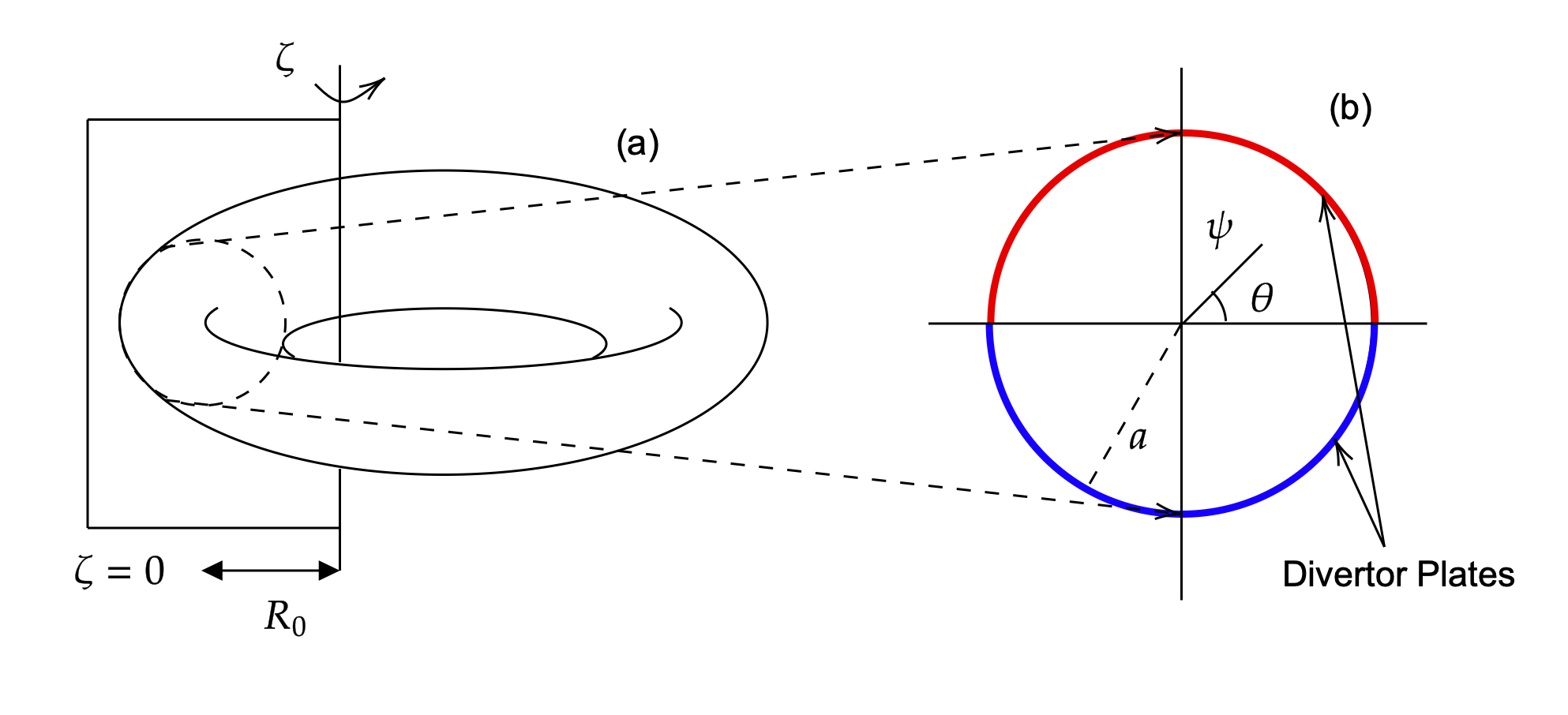}
    \caption{Schematic figure of (a) a Tokamak, (b) a Poincaré surface of section, showing the coordinates used to describe magnetic field lines and the representation of the divertor plates.}
    \label{fig:Toroidal_Space}
\end{figure}

Particles of mass $m$ and charge $q$, in a magnetized Tokamak plasma, tend to follow magnetic field lines, describing helical trajectories of gyro-radius $\frac{m v_p}{q B}$, where $B$ is the magnetic field intensity, and $v_p$ is the particle velocity projected on a plane orthogonal to the field direction. If the magnetic field is large enough, we can neglect, in a first approximation, the gyro-motion and assume that the plasma particles follow magnetic field lines. Within this approximation, we also neglect any other drift motions, derived from curvature and inhomogeneities of the magnetic field, as well as electric fields. Hence, the magnetic field structure gives a lower-order approximation of the charged particle motion.

Tokamaks are toroidal devices whose goal is the magnetic confinement of fusion plasmas. The Tokamak equilibrium magnetic field has two main components: a toroidal field $\mathbf{B}_T$ (created by coils external to the toroidal vessel) and the poloidal field $\mathbf{B}_P$ (due to the plasma current, which is created by induction). The magnetic field lines for $\mathbf{B} = \mathbf{B}_T + \mathbf{B}_P$ have a helical shape and lie on toroidal magnetic surfaces. The so-called magnetic axis is a degenerate magnetic surface.

In Fig.\ref{fig:Toroidal_Space}(a), we represent schematically a toroidal vessel, with minor radius $a$ and major radius $R_0$, with respect to its symmetry axis. The angle $\zeta$ is measured along the toroidal curvature, such that $0 \leq \zeta \leq 2\pi$. If we make a Poincaré surface of section at a plane $\zeta = 0$, the intersection of a magnetic field line can be described by polar coordinates $(r, \theta)$, where the origin is the intersection of the magnetic axis with this plane [\ref{fig:Toroidal_Space}(b)]. In this coordinate system, the magnetic surfaces are nested tori with radius $0 \leq r \leq a$ and $0 \leq \theta \leq 2\pi$. We characterize each magnetic surface by a quantity $\psi(r) = \frac{r^2}{a^2}$, in such a way that the magnetic axis and the plasma edge are located at $\psi = 0$ and $\psi = 1$, respectively.

It has been long known that the magnetic field line equations can be expressed as Hamilton’s equations, using $(\psi, \theta)$ as canonically conjugated variables, and where the angle $\zeta$ plays the role of time:
\begin{equation}
\frac{d\psi}{d\zeta} = -\frac{\partial H}{\partial \theta},
\end{equation}
\begin{equation}
\frac{d\theta}{d\zeta} = \frac{\partial H}{\partial \psi},
\end{equation}
where $H$ is the field line Hamiltonian. In an equilibrium Tokamak configuration, due to the axisymmetry, the Hamiltonian does not depend on $\zeta$, and $H(\psi, \theta)$ represents an integrable system. If the latter does not depend on $\theta$, Hamilton's equations are
\begin{equation}
\frac{d\psi}{d\zeta} = 0,
\end{equation}
\begin{equation}
\frac{d\theta}{d\zeta} = W(\psi),
\end{equation}
where $W$ is the so-called rotation number, which takes on a constant value on a given magnetic surface.

It turns out that $(\psi, \theta)$ are action and angle variables, respectively, such that magnetic surfaces are actually invariant tori of the field line Hamiltonian. Magnetic perturbations which break axisymmetry can be represented by a $\zeta$-dependent Hamiltonian $H_1$, in such a way that the perturbed field line configuration is described by
\begin{equation}
H(\psi, \theta, \zeta) = \int d\zeta \, W(\zeta) + H_1(\psi, \theta, \zeta).
\end{equation}

Let us denote by $(\psi_n, \theta_n)$ the coordinates of the $n$-th intersection of a magnetic field line with the plane $\zeta = 0$. We consider a Poincaré map $(\psi_n, \theta_n) \rightarrow (\psi_{n+1}, \theta_{n+1})$, relating the coordinates of two consecutive intersections of a field line. The solenoidal condition $\nabla \cdot \mathbf{B} = 0$ implies the conservation of the magnetic flux, and thus the Poincaré map must be area-preserving in the surface of section.

Balescu has proposed an analytical form for the Poincaré map, called Revtokamap\cite{balescu_tokamap_1998, balescu_hamiltonian_1998}, which is an area-preserving map used to model magnetic field line dynamics in toroidal plasmas exhibiting reversed magnetic shear, a configuration essential for suppressing turbulent transport in fusion devices \cite{Horton_Drift_1998, Levinton_Improved_1995, Terry_Suppression_2000}. The map is defined by the following iterative equations:

\begin{equation}
\begin{aligned}
    \psi_{n+1} &= \frac{1}{2} \left\{ P(\psi_n,\theta_n) + \sqrt{{[P(\psi_n,\theta_n)]}^2 + 4\psi_n} \right\}, \\
    \theta_{n+1} &= \theta_n + W(\psi_{n+1}) - \frac{K}{(2\pi)^2} \frac{\psi_{n+1}\cos(2\pi \theta_n)}{{(1+\psi_{n+1})}^2},
    \label{eq:revmap}
\end{aligned}
\end{equation}

where 
\begin{align}
    P(\psi_n,\theta_n) &= \psi_n - 1 - \frac{K}{2\pi} \sin(2\pi \theta_n), \\
    W(r) &= w \left[1 - a {(cr - 1)}^2 \right].
\end{align}

Here, $(r \in [0,1])$ represents the normalized radial coordinate (with $(r=1)$ marking the plasma edge), and $(\theta \in [0,1))$ is the poloidal angle. The perturbation strength $(K \in [0,2\pi))$ controls the external perturbation in magnetic field line trajectories, while $(W(r))$ governs the winding number profile. By choosing $(W(r))$ as a non-monotonic function, the map breaks the twist condition, creating transport barriers that mimic advanced tokamak scenarios \cite{haerter_basin_2023,souza_basin_2023,viana_transport_2021}. We have used as model parameters:

\begin{equation}
a = \frac{w-w_0}{w},\qquad c = 1+\left(\frac{w-w_1}{w-w_0}\right)^{1/2},
\end{equation}

with $w_0 = 0.3333$, $w = 0.6667$, $w_1 = 0.1667$.

These parameter values were chosen to replicate experimental conditions observed in the DIII-D tokamak experiment \cite{Greenfield_Transport_1997}. For all numerical simulations, we iterated $1024 \times 1024$ initial conditions uniformly distributed across both spatial directions, each evolved for $10^6$ iterations.

The area-preserving nature of the Revtokamap guarantees Hamiltonian dynamics, making it a computationally efficient alternative to solving complete magnetohydrodynamic equations. Its capability to accurately capture chaotic magnetic field line trajectories and reversed shear effects makes it a robust tool for investigating nonlinear particle transport in fusion plasmas, particularly under conditions involving reversed magnetic shear.

\section{Different Exit Configurations}
\label{sec:pri}

To explore how different configurations for opening the system influence the dynamical properties of particle escapes, we analyze three distinct exit configurations within the Revtokamap:  
\begin{itemize}
\item {\bf LR Case}. It was previously used in \cite{haerter_basin_2023}, and splits the phase space into two exits: $0 \leq \theta < 0.5$, labeled as $L$, and $0.5 \leq \theta \leq 1.0$, labeled as $R$.
\item {\bf BC Case}. Leveraging the periodicity of the angular coordinate $\theta$, a rotation of $90^\circ$ is applied, resulting in one basin at the boundaries: $0 \leq \theta < 0.25 \cup 0.75 < \theta \leq 1.0$, labeled as $B$, and another at the center: $0.25 \leq \theta \leq 0.75$, labeled as $C$.
\item {\bf Multi Case}. This division splits the phase space into ten different exits, allowing for a finer, more detailed analysis of escape dynamics.
\end{itemize}

\begin{figure*}[ht!]
    \centering
    \subfloat(a){\includegraphics[height=2.7in]{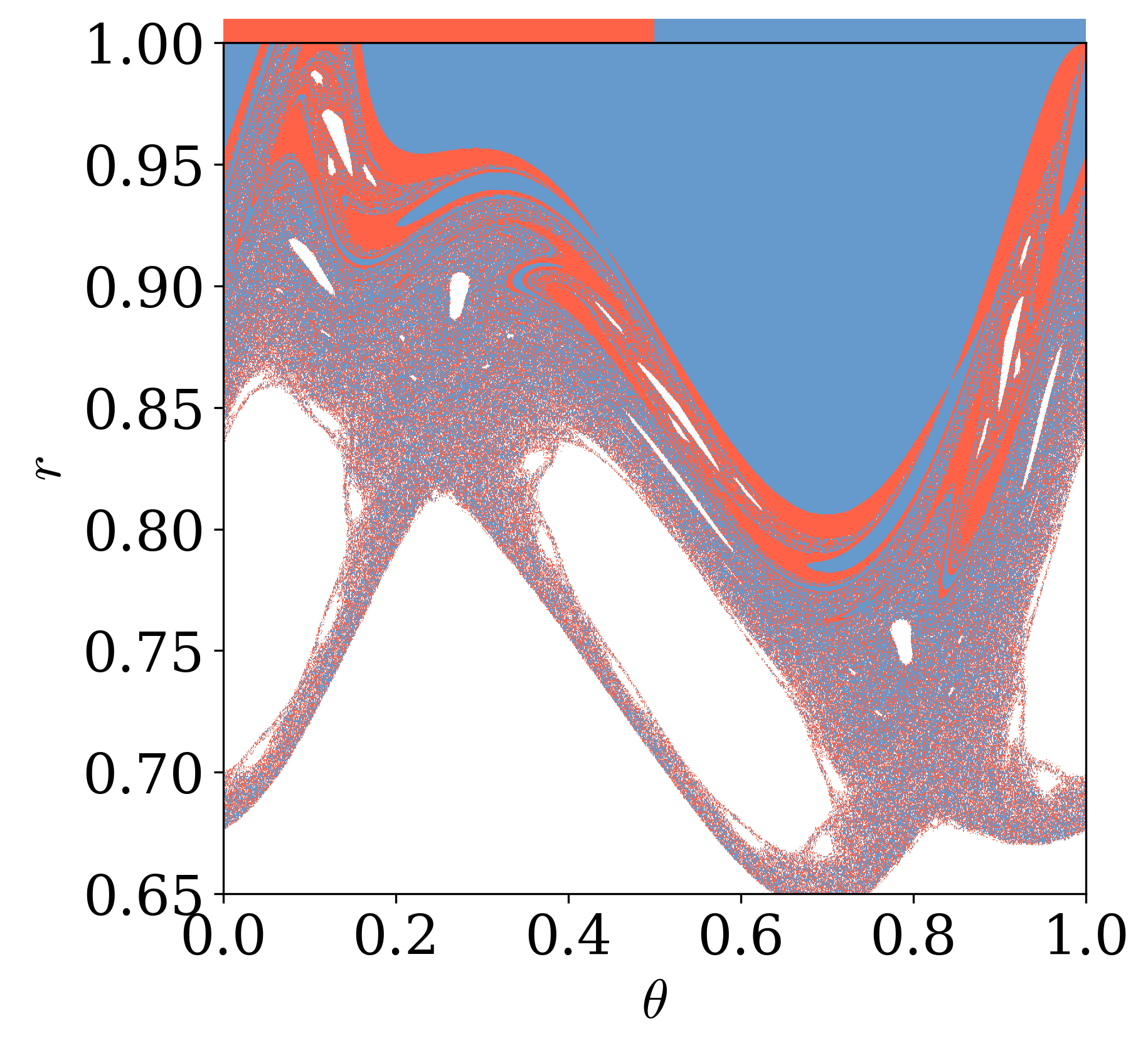}}
    \subfloat(b){\includegraphics[height=2.7in]{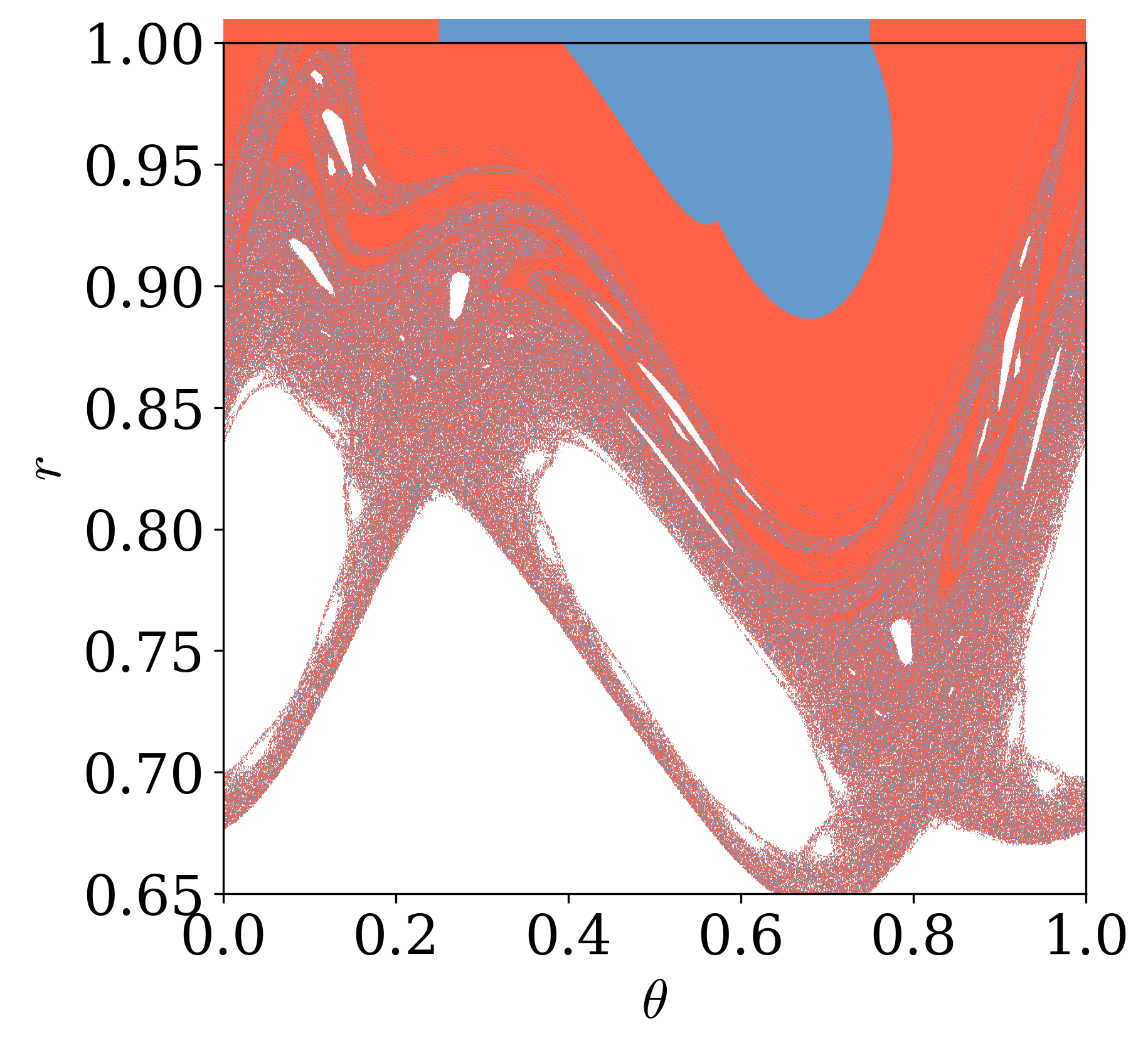}}
    
    \subfloat(c){\includegraphics[height=2.7in]{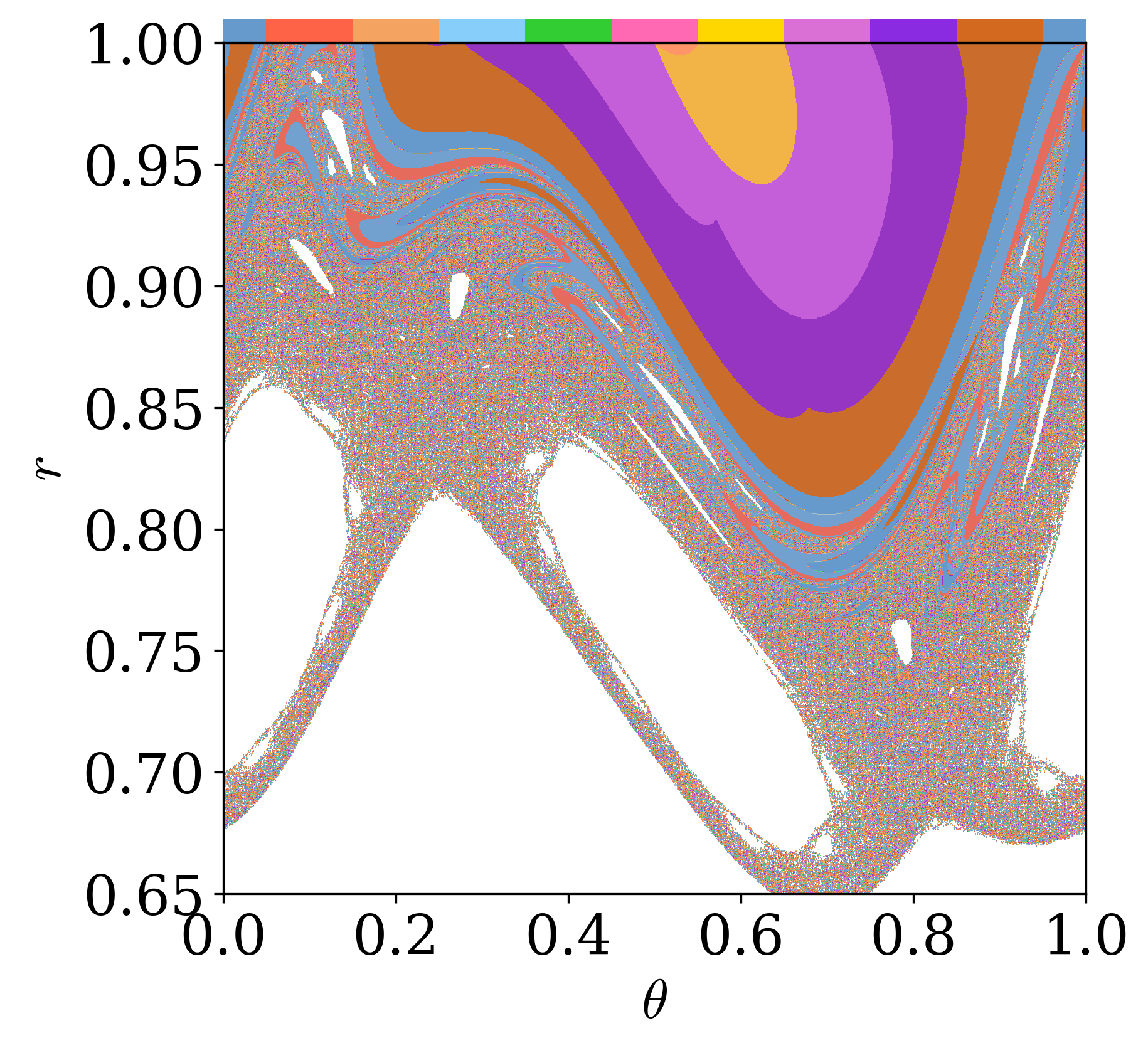}}
    \caption{Escape basins for $K=2.1$ in the three configurations analyzed: (a) $LR$ with the $L$ (red) and $R$ (blue) exits, (b) $BC$ with the $B$ (red) and $C$ (blue) exits, and (c) $Multi$ case, with $10$ different exits represented in the figure. The color bar above each plot indicates the corresponding exit region. These plots illustrate clearly how variations in the number and placement of exits alter the structure of escape basins, highlighting preferred regions of escape for this level of perturbation.}
    \label{fig:Basins}
\end{figure*}

Figure~\ref{fig:Basins} illustrates how different escape configurations affect the distribution of points within the escape basins, despite maintaining a constant perturbation value across all cases. In the $LR$ case, there is a notable intermixing of blue and red basin points near the quasiperiodic islands, demonstrating significant structural complexity and intertwined basin boundaries.

When the exits are rotated by $90^\circ$, the escape basin structure changes substantially, resulting in a clear predominance of the red basin over the blue one. In the $BC$ case, most particles escape along the symmetry axis, directly impacting the tokamak walls. In practical fusion scenarios, these regions would absorb greater energy fluxes and thus require enhanced protection measures. Although the central region also receives some chaotic particles, most escaping trajectories originate from the same regular region, identifiable as the distinct blue spot.

With this in mind, the $Multi$ case positions one escape basin along the symmetry axis and distributes the remaining nine basins around it, enabling a more detailed analysis of particle escape trajectories. This configuration provides deeper insights, despite the partially Wada nature of the basins previously identified in \cite{haerter_basin_2023}. A visual inspection of Fig.~\ref{fig:Basins}(c) indicates a high concentration of escaping particles in the blue basin (aligned with the symmetry axis) and its adjacent regions (red and dark brown), underscoring areas with preferential particle exits.

To quantify the effects of different exit configurations, we employ basin entropy analysis~\cite{daza_basin_2016}. This method partitions the escape region into a finite grid of  $N$ boxes, each of size $\varepsilon$, and calculates the corresponding escape probabilities, assuming equiprobability among all boxes:

\begin{equation}
  p_i = \frac{n_i}{\varepsilon^2},
\end{equation}

where $i \in [1, N_A]$ refers to each exit, and $N_A$ is the number of exits. The Shannon entropy for the $i$th cell is thus

\begin{equation}
    \label{entro}
    S_i = -\sum_{i=1}^{N_A} p_i \log p_i.
\end{equation}

Since this entropy is additive (extensive), the total entropy of the mesh is the sum over all boxes divided by their number:

\begin{equation}
S_b = \frac{1}{N} \sum_{i=1}^{N} S_i,
\label{eq:Sb}
\end{equation}

which is what we call basin entropy.

For boxes containing only one attractor, the Shannon entropy will be zero. Discarding these boxes in the calculation leads to another measure, the basin boundary entropy, where only boxes with more than one attractor are considered. It is defined as

\begin{equation}
  S_{bb} = \frac{1}{N_b} \sum_{i=1}^{N} S_i,
  \label{eq:Sbb}
\end{equation}

\noindent where $N_b$ represents the number of boxes with multiple attractors. This quantity is more sensitive to changes in the phase space and the presence of multiple attractors close to each other.

These two quantifiers can be utilized to analyze the complexity, mixture, and uncertainty of initial conditions within either escape basins or basins of attraction. They are closely related to classical metrics such as the fractal dimension of an attractor and the uncertainty exponent \cite{daza_basin_2016}. When $S_b \rightarrow 0$, almost all initial conditions within a small box of size $\varepsilon^2$ lead to the same exit, indicating minimal uncertainty and greater predictability of trajectories. Conversely, when $S_b \rightarrow \ln N_A$, initial conditions within the box are nearly equally distributed among all possible exits, corresponding to maximal uncertainty and unpredictability.

\begin{figure*}[ht!]
    \centering
    \subfloat{\includegraphics[height=2.7in]{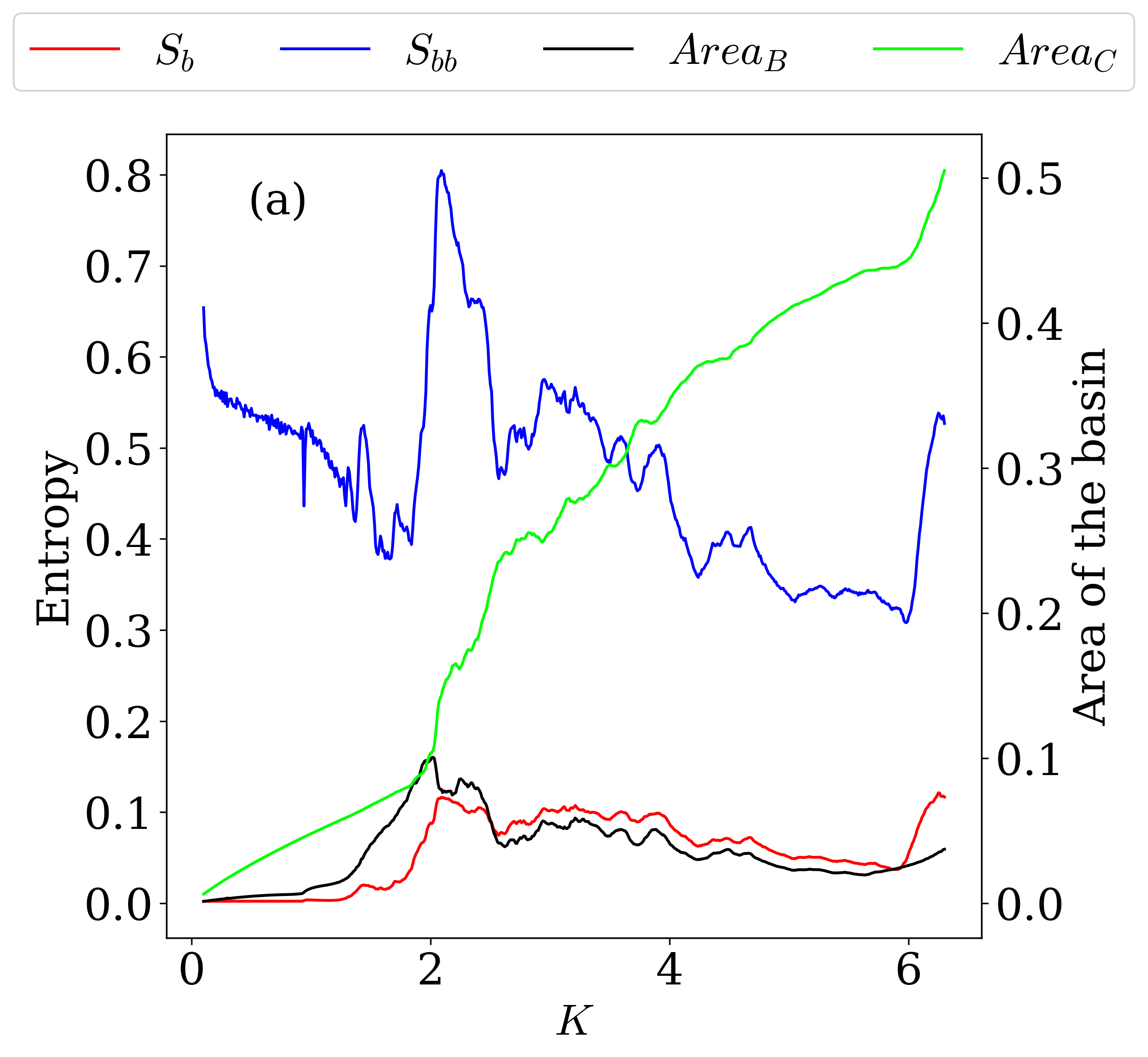}}
    \subfloat{\includegraphics[height=2.7in]{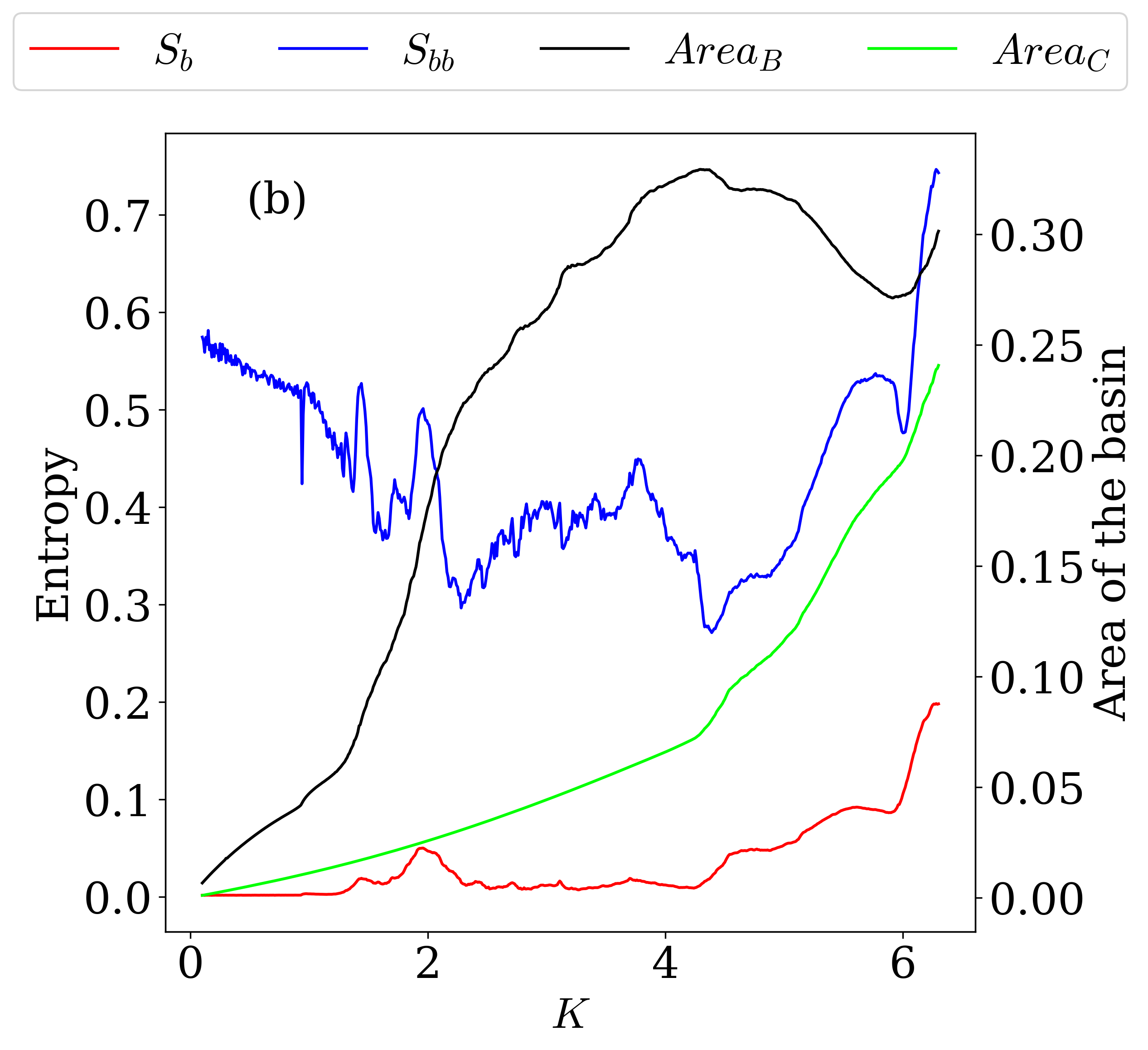}}
    
    \subfloat{\includegraphics[height=2.7in]{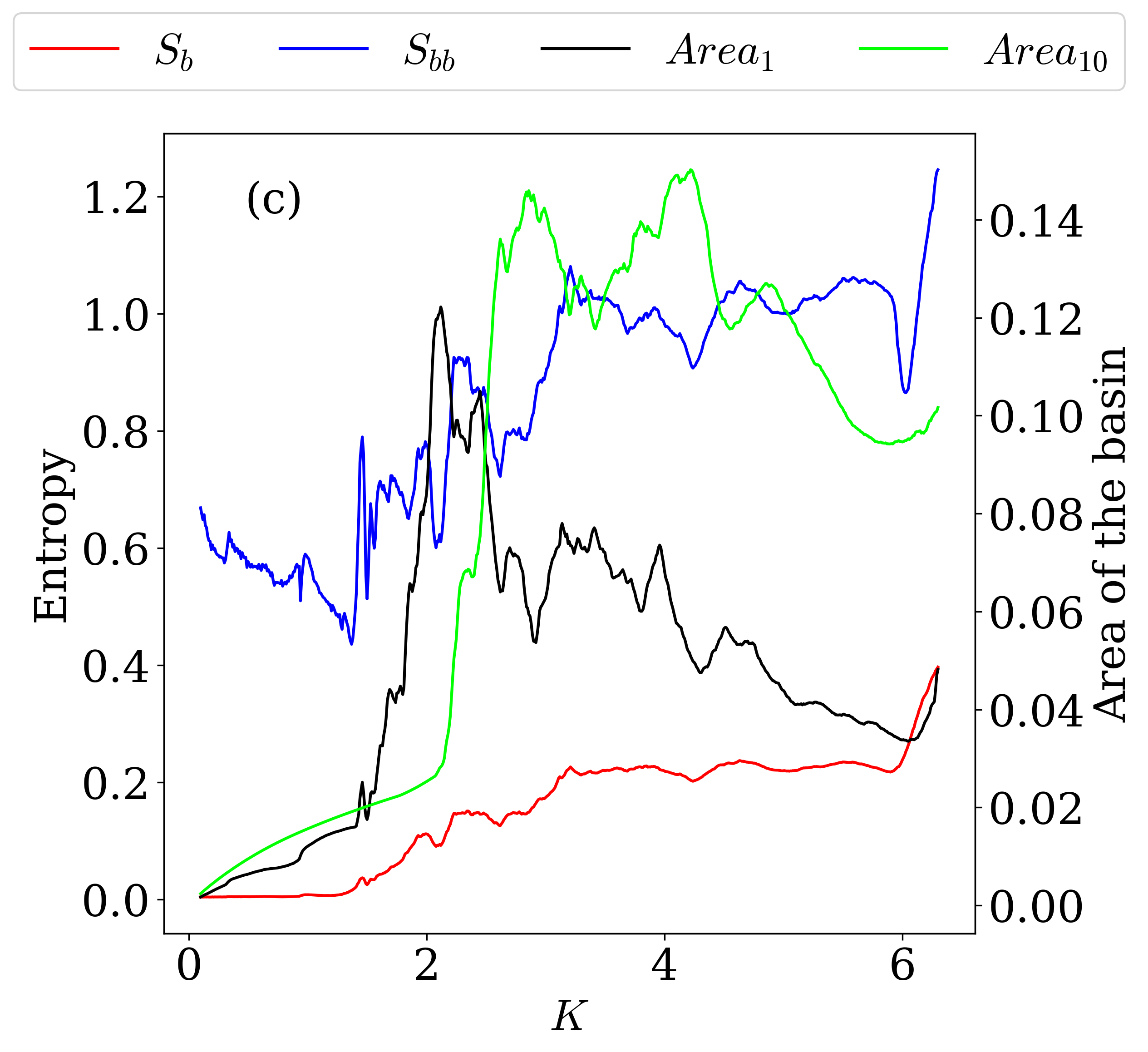}}
    \caption{Basin entropy (black), basin boundary entropy (green), and basin area fraction as a function of the parameter $K$ for the Revtokamap under the three analyzed configurations:  (a) $LR$ case, (b) $BC$ case, and (c) $Multi$ case. These plots illustrate the relationships among these measures, highlighting how basin entropy and basin boundary entropy are influenced by changes in occupied basin areas and how all these quantities evolve with increasing perturbation, reflecting greater chaoticity in the system.}
    \label{fig:entropies}
\end{figure*}

Figure~\ref{fig:entropies} displays the basin entropy, basin boundary entropy, and the occupied basin area for each of the three cases analyzed. As expected, the results across all configurations show that basin entropy increases with larger values of the perturbation parameter. This rise occurs due to enhanced chaotic behavior in the system, leading to greater fractality and complexity of basin structures.

An important observation left unexplained in previous work \cite{haerter_basin_2023} is the abrupt transition observed in both basin entropy and basin boundary entropy around the critical perturbation value $K_c = K \approx 2$. This phenomenon is clearly reproduced in our analysis, as illustrated in Fig.~\ref{fig:entropies}(a). Specifically, basin entropy initially follows a smooth, gradually increasing trend at lower perturbation values until reaching the critical point $K_c$. At this threshold, we observe a sharp, pronounced increase in the basin entropy, indicating a sudden structural transition in the escape basin topology. Interestingly, beyond this critical value, basin entropy begins to decrease, suggesting a reorganization or simplification in basin structures as the perturbation further increases. This complex dynamical behavior likely arises from a significant restructuring of chaotic regions and their boundaries, driven by changes in the stability and interactions among invariant sets within phase space.

In Fig.~\ref{fig:entropies}(a), it is evident that near $K_c$, the area of the $R$ basin increases abruptly, while the $L$ basin area remains almost constant. This has a direct effect on $S_b$ and $S_{bb}$, with the highest basin boundary entropy occurring when both areas reach their maximum values. The dominance of one exit over the other, despite the system exhibiting more chaotic behavior due to the increasing perturbation parameter, leads to a decrease in basin entropy.

On the other hand, when the system is rotated, as shown in Fig.~\ref{fig:Basins}(b), the basin entropy remains close to zero for higher values of $K$, with the area of the red basin dominating over the blue one. This persists until a high perturbation value of $K \approx 4$, where the blue basin area begins to increase, and with it, the basin entropy starts to rise. This behavior indicates a preference for escape positions along the symmetry axis, which is only altered at high perturbation values.

The $Multi$ case provides further insight into this phenomenon. Due to the arbitrary choice of exits, certain preferences can be established, influencing the interpretation of the results. By positioning one of the exits on the symmetry axis and increasing the number of exits, we can observe a more precise behavior of the basins and how they evolve with the perturbation parameter.

From Fig.~\ref{fig:Basins}(c), a relative tendency of particles to move to the right is evident, as all smooth regions are connected to the exits on the right side of the polar direction. Additionally, when analyzing the basin entropy and areas in Fig.~\ref{fig:entropies}(c), a counterclockwise movement of the basin areas is observed. For lower values of $K$, there is a higher escape rate through the first exit, located near the symmetry axis. Upon reaching $K \approx 2.5$, this rate decreases, and simultaneously, the area of the adjacent basin to the left begins to increase at a similar rate, indicating a shift in the escape orbits in that direction.

\section{Conclusions}
\label{sec:concl}
In this study, we have examined how different methods of opening Hamiltonian systems influence the dynamical properties of escape trajectories, using the Revtokamap as our primary model. This choice was motivated by the Revtokamap’s accurate representation of magnetic field lines in toroidally confined plasmas, where understanding particle escape mechanisms is critical for improving plasma confinement and reactor performance. The implementation of linear interpolation to precisely determine exit points enabled a more accurate classification of escape basins and provided deeper insights into the dynamics governing the movement and mixing of escaping orbits. This methodological approach facilitated the identification of structural transitions and highlighted how basin complexity evolves under varying escape configurations.

By systematically varying the escape configuration across three distinct cases, we were able to observe and characterize both quantitatively and qualitatively how each choice of exit impacts escape dynamics. Visual comparisons of escape basin distributions clearly revealed a preferential escape through a specific region aligned along the symmetry axis. This observation was further supported by detailed analysis of basin area fractions, basin entropy, and basin boundary entropy in each configuration, confirming that escape behavior and basin complexity are strongly influenced by the strategic placement and number of exits.

The simplest and most intuitive approach to partitioning phase space, without explicitly considering its underlying dynamical properties, resulted in abrupt discontinuities in basin entropy at certain perturbation values. This outcome underscores the critical importance of incorporating the inherent nonlinear dynamics of the system when defining escape regions. Designing escape regions based solely on spatial convenience or intuitive symmetry can significantly influence the resulting complexity and predictability of basin structures, highlighting the need for dynamically informed strategies in the context of fusion device optimization.

The shifting trajectories of escaping orbits create problematic intersections between adjacent exit regions, resulting in abrupt transitions and increased complexity along basin boundaries. These intersections disrupt the smoothness of the basin boundaries, causing sudden and pronounced increases in basin mixing. Consequently, as exits begin to overlap dynamically, there are sharp changes in both basin entropy and basin boundary entropy, reflecting the enhanced uncertainty and complexity introduced by these crossing regions.

When the exits were rotated, the uncertainty in basin structures was significantly suppressed and only became prominent at higher perturbation values $K$. This result indicates a more stable and favorable exit configuration, particularly relevant for identifying regions in real fusion devices that would benefit from reinforced plasma-facing components.

Additionally, by increasing the number of exits, we obtained a more detailed and precise characterization of escape basin structures. This finer granularity clearly confirmed the presence of preferential escape pathways, highlighting specific regions where escaping particles are more concentrated and thus guiding optimized design and protection strategies for tokamak walls.

\section*{Acknowledgments}

This work has been supported by grants from the Brazilian Government Agencies CNPq and CAPES. P. Haerter received partial financial support from the following Brazilian government agencies: CNPq (140920/2022-6), CAPES (88887.898818/2023-00).  R. L. Viana received partial financial support from the following Brazilian government agencies: CNPq (403120/2021-7, 301019/2019-3), CAPES (88881.143103/2017-01). Finally, M.A.F. Sanjuan acknowledges financial support by the Spanish State Research Agency (AEI) and the European Regional Development Fund (ERDF, EU) under Projects No.~PID2019-105554GB-I0 (MCIN/AEI/10.13039/501100011033) and No.~PID2023-148160NB-I00 (MCIN/AEI/10.13039/ 501100011033).

\bibliographystyle{elsarticle-num}
\bibliography{Refs}

\end{document}